\documentclass[showpacs,preprintnumbers,amsmath,amssymb,prl,twocolumn,
superscriptaddress,footinbib]{revtex4}
\usepackage{graphicx}
\usepackage{bm}
\usepackage{subfigure}
 \newcommand{\BF}[1]{\mbox{\boldmath $#1$}}

\def\varphib{{\BF \varphi}}
\def\psib{{\BF \psi}}

\begin{document}

\title{Quantum critical scaling behavior of deconfined spinons}

\author{F. S. Nogueira}
\affiliation{Institut f\"ur Theoretische Physik, Freie Universit{\"a}t Berlin,
Arnimallee 14, D-14195 Berlin, Germany}

\author{S. Kragset}
\affiliation{Department of Physics, Norwegian University of
Science and Technology, N-7491 Trondheim, Norway}

\author{A. Sudb{\o}}
\affiliation{Department of Physics, Norwegian University of
Science and Technology, N-7491 Trondheim, Norway}

\date{Received \today}

\begin{abstract}
We perform a renormalization group analysis of some important effective 
field theoretic models for deconfined spinons. 
We show that deconfined spinons are critical for an isotropic SU(N) Heisenberg antiferromagnet, 
if $N$ is large enough. We argue that nonperturbatively 
this result should persist down to $N=2$ and provide further evidence for 
the so called deconfined quantum criticality scenario. 
Deconfined spinons are also shown to be critical for the case describing 
a transition between quantum spin nematic and dimerized phases. 
On the other hand, 
the deconfined quantum criticality 
scenario is shown to fail for a class of easy-plane models.  
For the cases where deconfined 
quantum criticality occurs, we calculate the critical exponent $\eta$ for the decay of 
the two-spin correlation function to first-order in $\epsilon=4-d$. We also note 
the scaling relation 
$\eta=d+2(1-\varphi/\nu)$ connecting the exponent $\eta$ for the decay 
to the correlation length exponent $\nu$ and the crossover exponent $\varphi$.         
\end{abstract} 

\pacs{75.30.Kz,64.60.Cn,71.30.+h,}

\maketitle 

The most remarkable incarnation of the Landau-Ginzburg theory of 
phase transitions is the one embodied by Wilson's renormalization group (RG)  
\cite{Wilson}. According to this point of view, the Landau-Ginzburg theory 
is uniquely determined by the effective coupling constants obtained 
by integrating out high-energy modes. In this way, the large distance 
scaling behavior of different physical quantities  
is governed by the fixed points in the space of 
coupling constants. This is the so called Landau-Ginzburg-Wilson (LGW)  
paradigm of phase transitions \cite{Senthil}. The LGW paradigm is known 
to fail in a number of quantum phase transitions. One prominent example is 
the transition between the N\'eel state to a valence bond solid (VBS) state 
in a two-dimensional Mott insulator \cite{RS}. 
This transition features a quantum critical 
point (QCP), which is at odds with the LGW scenario that would predict a first-order 
phase transition. The crucial observation in this context is that both phases 
break symmetries in distinct spaces: the N\'eel state breaks the SU(2) symmetry of 
the Hamiltonian, while the paramagnetic VBS state breaks lattice symmetries. 
A {\it continuous} such order-order phase transition 
would not be captured by a LGW-like point of view \cite{Chakra}.     

For an SU(2) Heisenberg antiferromagnet 
the spinons $z_\alpha$ are the elementary constituents of the spin orientation field 
${\bf n}$. We have  
$n_a={\bf z}^\dagger\sigma_a{\bf z}$, $a=1,2,3$, where ${\bf z}=(z_1,z_2)$ and 
$\sigma_a$ are the Pauli matrices.  
This is the so called CP$^{1}$ representation of 
the SU(2) spins. 
There is an inherent {\it local} gauge invariance in this representation, since 
${\bf n}$ remains invariant when the spinon fields change by a local phase factor, 
i.e., $z_\alpha\to e^{i\theta(x)}z_\alpha$. Thus, it seems to be  natural to 
effectively describe a Mott insulator through a gauge theory coupled to ``spinon matter''. 
The gauge field here is an {\it emergent 
photon}: it is dynamically generated as a 
consequence of the local gauge invariance of ${\bf n}$ 
in terms of the spinon fields. 
Note that only 
expectation values of gauge-invariant operators can be nonzero, in agreement with 
Elitzur's theorem \cite{Elitzur}. The VBS order parameter is also a gauge-invariant 
expectation value, since it is proportional to $\langle{\bf n}_i\cdot{\bf n}_j\rangle$, 
with $i$ and $j$ being nearest neighbor sites in a square lattice. 
The spinons are 
confined in both the N\'eel and VBS phases. Indeed, spinon deconfinement would make ${\bf n}$ 
fall apart leading to a vanishing of both spin and VBS order parameters. A point 
in the phase diagram  where the spinons may deconfine is at a QCP, where both 
order parameters are supposed to vanish. 
To actually demonstrate that this happens, is not an easy task. 
The main 
argument \cite{Senthil} behind the concept of deconfined quantum criticality (DQC) is a 
topological one:         
spinon deconfinement occurs due to a destructive 
interference mechanism between the instantons and the Berry phase \cite{Senthil}. 
This mechanism 
was recently observed numerically \cite{Kragset} 
for the case of an easy-plane antiferromagnet.  
However, in this case the instanton cancellation mechanism leads actually to 
a weak first-order phase transition \cite{Kragset}.
In Ref.  
\cite{Kuklov_2006} a first-order phase transition in an easy-plane model of 
votex loops was also found, but there Berry phase effects were not considered. 

One of the most important aspects of DQC is the large value of the critical exponent 
$\eta$ for the decay of the correlation function 
${\cal G}(x)\equiv\langle{\bf n}(x)\cdot{\bf n}(0)\rangle$ as compared with the value obtained 
through the LGW approach. This correlation function is highly relevant experimentally. Therefore, 
it is important to be able to calculate $\eta$ in a systematic way. The exponent $\eta$ has been 
calculated in Monte Carlo simulations for two-dimensional  
Heisenberg antiferromagnets with instanton suppression \cite{Motrunich} and with 
four-spin interactions \cite{Sandvik}. The obtained results are $\eta\approx 0.7$ and 
$\eta\approx 0.26$, respectively. 

One of the main results of the present paper will be 
the calculation  of $\eta$ in first-order in $\epsilon=4-d$, where $d$ is the dimension of space-time. 
This will be done for two different DQC regimes: (i) the N\'eel-VBS 
transition \cite{Senthil} and (ii) the phase transition between quantum spin nematic and dimerized 
phases \cite{Grover}.

It was argued in Ref. \cite{Senthil} that  
for certain isotropic $SU(N)$ 
symmetric Heisenberg antiferromagnets the quantum critical point is governed by 
the euclidean Lagrangian
\begin{eqnarray}
\label{S-linear}
{\cal L}&=&\frac{1}{2}(\epsilon_{\mu\nu\lambda}\partial_\nu A_\lambda)^2+
\sum_{\alpha=1}^N|(\partial_\mu-ie_0A_\mu)z_\alpha|^2\nonumber\\
&+&r_0\sum_{\alpha=1}^N|z_\alpha|^2+\frac{u_0}{2}
\left(\sum_{\alpha=1}^N|z_\alpha|^2\right)^2,
\end{eqnarray}
where the parameter $r_0$ is tuned in such a way that the system is at its 
critical point. The above Lagrangian corresponds to an Abelian Higgs model in 
euclidean space with an $O(2N)$ global symmetry. It can also be thought of as the 
free energy of a Ginzburg-Landau (GL) model with $N$ complex order parameter 
fields.    
In this case the upper critical dimension is four. Thus, the RG analysis 
should be made in $d=4-\epsilon$ dimensions. We define dimensionless couplings 
$g=\mu^{-\epsilon}u_r$ and $f=\mu^{-\epsilon}e_r^2$, where $u_r$ and $e_r$ are 
the renormalized counterparts of $u_0$ and $e_0$, respectively. The 
RG $\beta$ functions $\beta_f\equiv\mu\partial f/\partial\mu$ and 
$\beta_g\equiv\mu\partial g/\partial\mu$ are straightforwardly obtained 
employing standard techniques \cite{KSF}: 
\begin{equation}
\label{betaf}
\beta_f=-\epsilon f+\frac{N}{3}f^2,
\end{equation}
\begin{equation}
\label{betag}
\beta_g=-\epsilon g-6fg+(N+4)g^2+6f^2.
\end{equation}
The above RG equations are well known 
 in the context of the GL model  \cite{Halperin-Lubensky-Ma}.  
An infrared stable fixed point with $f\neq 0$ is found 
only for $N$ large enough, namely, $N>182.9$. Unless $N$ is greater 
than this value, no second-order phase transition is predicted 
by this RG analysis. In the past this result led to the conclusion 
\cite{Halperin-Lubensky-Ma} that thermal fluctuations 
turn the phase transition in a superconductor into a first-order 
one, since there the actual number of components is $N=1$. 
It did not take too long to realize that this result is incorrect 
\cite{Dasgupta,Kleinert-tric}. Actually the large, but finite, $N$ 
result reflects the strong-coupling features of the $N=1$ theory,  
which cannot be captured by the RG analysis in $d=4-\epsilon$ 
dimensions. {\it This does 
not mean that the first-order 
transition cannot occur}. It turns out that the complete phase 
diagram features a tricritical point \cite{Kleinert-tric} at a 
value of the Ginzburg parameter $\kappa=u/\sqrt{2}e$ smaller than 
the mean-field GL value separating the type I from the type II 
regimes, i.e., $\kappa=1/\sqrt{2}$. Earlier calculations based on 
duality arguments \cite{Kleinert-tric} give the value 
$\kappa_t\approx 0.8/\sqrt{2}$, a result recently confirmed by large scale 
Monte Carlo simulations \cite{Sudbo-tric}. The weak-coupling regime 
at low $N$ 
captured by the RG functions (\ref{betaf}) and (\ref{betag}) 
corresponds to the one in which $\kappa\ll\kappa_t$. 
An RG analysis in fixed dimension $d=3$ \cite{HT-1}, though less well controlled 
than the one near four dimensions, indicates that the critical value of $N$ can be 
drastically reduced if a resummed higher order calculation is performed 
\cite{HT-2}. Other interesting effects arise as 
the number of components gets large enough for the case where the 
gauge field is compact. For example, there is a recent numerical evidence 
for a Coulomb-like phase in a compact abelian gauge theory coupled 
to multiflavor CP$^1$ fields \cite{Ichinose}.     

We can make an analysis of deconfined quantum criticality 
that parallels the case of superconductors in the 
neighborhood of the critical temperature. In contrast to the 
superconductor, in this case the physical value of the parameter 
$N$ is given by $N=2$. Hence, we 
have two classes of (meron) vortices \cite{Senthil,Babaev}. The 
above discussion in the context of superconductors 
shows that in principle we can also have 
{\it deconfined quantum tricriticality}.            

The correlation function  
${\cal G}(x)=\langle{\bf n}(x)\cdot{\bf n}(0)\rangle$  
in the CP$^{N-1}$ representation reads 
\begin{eqnarray}
\label{G}
{\cal G}(x)&=&2\langle{\bf z}^*(x)\cdot{\bf z}(0)~{\bf z}(x)\cdot{\bf z}^*(0)\rangle
\nonumber\\
&-&\frac{2}{N}\langle|{\bf z}(x)|^2|{\bf z}(0)|^2\rangle.
\end{eqnarray}
%
At the deconfined 
QCP this correlation function scales as ${\cal G}(x)\sim 1/|x|^{d-2+\eta}$ \cite{Note}. In 
order to compute $\eta$, let us analyse the scaling behavior of the two 
four-spinon correlation functions in Eq. (\ref{G}).  
 
The scaling behavior of $\langle|{\bf z}(x)|^2|{\bf z}(0)|^2\rangle$ is obtained 
by considering the scaling dimension of the operator $|{\bf z}(x)|^2$. This is 
easily obtained by performing derivatives with respect to $r_0$ of the 
logarithm of the functional integral and doing dimensional analysis. The result is 
a scaling behavior of the form
$\langle|{\bf z}(x)|^2|{\bf z}(0)|^2\rangle\sim 1/|x|^{d-2+\eta_4}$, where 
$\eta_4=d+2(1-1/\nu)$, with $\nu$ being the correlation length exponent. This 
leads to a vanishing of those correlations in momentum space as $p\to0$, except 
for the mean-field case where $\eta_4=d-2$. 
Indeed, beyond mean-field theory we have $\nu>2/d$ and thus it is clear that $\eta_4>d-2$ when the 
fluctuations are included. This result is important because it legitimates the softening 
of the CP$^{N-1}$ constraint $|{\bf z}|^2=1$. In the critical regime we can simply neglect 
the second term in Eq. (\ref{G}). 

Let us consider now the scaling behavior of 
$\langle{\bf z}^*(x)\cdot{\bf z}(0)~{\bf z}(x)\cdot{\bf z}^*(0)\rangle$. This 
correlation function is associated with a mass anisotropy term, which is obviously 
not generated by quantum fluctuations in a SU(N) theory like the one in Eq. (\ref{S-linear}). 
However, we can consider it as a source term and compute the so called 
crossover exponent $\varphi$ \cite{ZJ}. The exponent $\eta$ is then obtained 
by replacing $1/\nu$ in the expression for $\eta_4$ by $\varphi/\nu$. Therefore, 
$\langle{\bf z}^*(x)\cdot{\bf z}(0)~{\bf z}(x)\cdot{\bf z}^*(0)\rangle\sim 1/
|x|^{d-2+\eta}$, where $\eta=d+2(1-\varphi/\nu)$. This result is obtained as follows. 

The anomalous dimensions of all quadratic operators, leading to mass anisotropy or not, 
can be derived from a ``matrix exponent'' 
$\eta^{(2)}_{\alpha\beta,\gamma\delta}=\lim_{\mu\to 0}\mu\partial\ln[Z^{(2)}_{\alpha\beta,\mu\nu}
(Z^{-1/2})_{\mu\gamma}(Z^{-1/2})_{\nu\delta}]/\partial\mu$ (here a summation over 
repeated greek indices is implied), where $Z^{(2)}_{\alpha\beta,\mu\nu}$ is the renormalization 
associated to the insertion of a quadratic operator and $Z_{\alpha\beta}$ the spinon 
wave function renormalization \cite{ZJ}. From the eigenvalues of this  
matrix exponent we can determine both $\nu$ and $\varphi$ or, equivalenty, $\eta_4$ and $\eta$. 
These eigenvalues are the anomalous dimensions of the composite operators 
$|{\bf z}|^2$ and $z_\alpha^*z_\beta$, with $\alpha\neq\beta$. We will call these anomalous 
dimensions $\eta_2$ and $\eta_2'$, respectively. Now it is straightforward to use dimensional analysis 
to obtain that $\eta_4=d-2-2\eta_2=d+2(1-1/\nu)$ and $\eta=d-2-2\eta_2'=d+2(1-\varphi/\nu)$. 
 
For the DQC regime described by 
Eq. (\ref{S-linear}) we have at one-loop order
\begin{equation}
\eta^{(2)}_{\alpha\beta,\gamma\delta}=-NP_{\alpha\beta,\gamma\delta}g_*+(3f_*-g_*)I_{\alpha\beta,\gamma\delta},
\end{equation}
where $I_{\alpha\beta,\gamma\delta}=(\delta_{\alpha\gamma}\delta_{\beta\delta}
+\delta_{\alpha\delta}\delta_{\beta\gamma})/2$, 
$P_{\alpha\beta,\gamma\delta}=\delta_{\alpha\beta}\delta_{\gamma\delta}/N$, and 
$g_*$ and $f_*$ are the infrared stable fixed points associated with the $\beta$ functions 
(\ref{betag}) and (\ref{betaf}). The eigenvalue $\eta_2$ corresponding to the 
eigenvector $\delta_{\gamma\delta}$ determines the critical exponent 
$\nu$ as $1/\nu=2+\eta_2$. The second eigenvalue, $\eta_2'$ determines 
$\varphi$ through $\eta_2'=\varphi/\nu-2$. Explicitly, we have 
$\eta_2=-(N+1)g_*+3f_*$ and $\eta_2'=-g_*+3f_*$. Therefore, we obtain 
to order $\epsilon$ and for $N>182.9$ the result
\begin{equation}
\eta=2-\left(1+\frac{18}{N}\right)\epsilon+2g_*,
\end{equation}
where $g_*=(18+N+\sqrt{N^2-180N-540})\epsilon/[2N(N+4)]$. Using the lowest value of 
$N$ for which the stable fixed point exists (i.e., $N=183$), we obtain after setting $\epsilon=1$ 
the result $\eta=609/671\approx 0.9076$. The LGW result to order $\epsilon$ would 
give $\eta=0$ and a small correction to order $\epsilon^2$. Note that the local 
gauge invariance is essential in order to get a value smaller than one for $\eta$. If 
we consider the model without any gauge coupling, we obtain $\eta=2-[1-2/(N+4)]\epsilon$, 
which is an expression valid for all values of $N$, since in this case the perturbative 
fixed point exists even for $N=1$. This leads for $N=183$ and $\epsilon=1$ to the result 
$\eta=189/187>1$.     

Now we consider the scaling behavior of the spin $S=1$ Hamiltonian \cite{Harada,Grover}
\begin{equation}
H=\sum_{\langle i,j\rangle}\left[J~{\bf S}_i\cdot{\bf S}_j-K({\bf S}_i\cdot{\bf S}_j)^2\right],
\end{equation}
where both nearest-neighbor couplings $J$ and $K$ are positive. This Hamiltonian 
describes the phase transition between a quantum spin nematic phase and a dimerized phase. 
Recent numerical results indicate that this model exhibits a second-order phase transition 
if the ratio $K/J$ is large enough. As pointed out in Ref. \cite{Grover}, the LGW paradigm 
would in this case predict a first-order phase transition, at odds with the results observed 
numerically \cite{Harada}. Here we will show using the RG that in this model a 
second-order phase transition occurs for large enough $N$.  
The field theory of the above model was derived 
recently \cite{Grover} and is given by the Lagrangian
\begin{eqnarray}
\label{nematics}
{\cal L}&=&\frac{1}{2}(\epsilon_{\mu\nu\lambda}\partial_\nu A_\lambda)^2+
|(\partial_\mu-ie_0A_\mu){\bf D}|^2+r_0|{\bf D}|^2\nonumber\\
&+&\frac{u_0+v_0}{2}(|{\bf D}|^2)^2-\frac{v_0}{2}({\bf D})^2({\bf D}^*)^2,
\end{eqnarray}
where $v_0>0$ and ${\bf D}$ is a complex vector with three components.   

To see how the second-order transition emerges, let us  
write $D_i=(\varphi_i+i\psi_i)/\sqrt{2}$, with $i=1,2,3$. The local interaction 
between the scalar fields become
\begin{equation}
{\cal L}_{\rm int}=\frac{u_0}{8}(\varphib^2+\psib^2)^2
-\frac{v_0}{2}(\varphib\cdot\psib)^2.
\end{equation}
The above equation features an interaction reminiscent of certain classical models for frustated 
magnetism \cite{Delamotte}. 
In order to perform the RG analysis, we will consider a generalization of the 
model such that $\varphib$ and $\psib$ have each $N$ components, with the 
physically relevant case corresponding to $N=3$. The $\beta$ function for 
the gauge coupling is given once more by Eq. (\ref{betaf}). By introducing  
dimensionless couplings $g=\mu^{-\epsilon}u$ and $h=\mu^{-\epsilon}v$ we obtain 
the one-loop $\beta$ functions:
\begin{equation}
\beta_g=-\epsilon g-6fg+(N+4)g^2+2h^2-2gh+6f^2,
\end{equation}
\begin{equation}
\beta_h=-\epsilon h-6fh-(N+2)h^2+6gh.
\end{equation}
It is useful to analyse first the case where $f=0$. The physically meaningful  
case corresponds to fixed points where $h\geq 0$. The relevant fixed point 
in this case has coordinates $g_*=2\epsilon/(N^2+8)$ and $h_*=(2-N)\epsilon/(N^2+8)$. This 
fixed point is stable only for $N=3$, but then we would have $h_*<0$, which is incompatible 
with the physical constraints of the model \cite{Grover}. 

Remarkably, for $f_*=3\epsilon/N$ a stable fixed point is found for $N>232.98$:
\begin{equation}
\label{g-fp-nem}
g_*=\frac{360-12N+2N^2+N^3+(N+2)\sqrt{\Delta}}{2N(64+8N+8N^2+N^3)}~\epsilon,
\end{equation}
\begin{equation}
h_*=\frac{3\sqrt{\Delta}-36-104N-21N^2-N^3}{N(64+8N+8N^2+N^3)}~\epsilon,
\end{equation}
where $\Delta=N^4-224N^3-2072N^2-4608N-22896$. Once more, just like in the 
isotropic case, we interpret the existence of stable fixed points at large values of $N$ as 
a strong evidence of deconfined quantum criticality. The exponent $\eta$ is calculated similarly as 
before, giving the result $\eta\approx 0.927$ for $N=233$ and $\epsilon=1$. 

Next, we consider an easy-plane version of the model (\ref{S-linear}). 
This amounts to adding an interaction term of the form $v_0(|z_1|^2-|z_2|^2)^2/2$. 
Previous results on the easy-plane model \cite{Senthil,Motrunich} indicated 
that a second-order phase transition would occur. This conclusion was based on 
the analysis of the {\it deep} easy-plane limit of the model. This regime 
corresponds to a large $v_0$ such that $|z_1|^2\approx|z_2|^2$. However, recent 
Monte Carlo simulations \cite{Kuklov_2006,Kragset} performed in this regime showed that the 
transition is actually (weakly) first-order. 

In order to facilitate the RG calculations it is convenient to write the complete 
local interaction between the spinons in the following form:
\begin{equation}
\label{aniscp1}
{\cal L}_{\rm int}=\frac{\bar u_0}{2}(|z_1|^4+|z_2|^4)
+w_0|z_1|^2|z_2|^2,
\end{equation}
where $\bar u_0=u_0+v_0$ and $w_0=u_0-v_0$. Let us introduce the 
renormalized dimensionless couplings $g=\bar u\mu^{-\varepsilon}$ 
and $h=w\mu^{-\varepsilon}$, where $\bar u$ and $w$ are the 
renormalized counterparts of $\bar u_0$ and $w_0$, respectively. In order 
to have the same total number of complex components as before, we will consider 
$N/2$ components of $z_1$ and $z_2$, with $N$ even. By this we 
mean a rewriting of the interaction, such that the system has a 
$O(N)\times O(N)$ symmetry. The one-loop 
$\beta$ function for the gauge 
coupling is the same as before. The other $\beta$ functions are 
%
\begin{equation}
\label{betagbar}
\beta_{\bar g}= -\epsilon g-6gf+\frac{N+8}{2}g^2+
\frac{N}{2}h^2+6f^2,
\end{equation}
\begin{equation}
\label{betah}
\beta_h=-\epsilon h-6hf+(N+2)gh+2h^2+6f^2.
\end{equation}
It is instructive to consider first the model for vanishing gauge coupling ($f=0$). 
In this case besides the Gaussian ($g_*=h_*=0$) and Heisenberg [$g_*=2\epsilon/(N+8)$ 
and $h_*=0$] fixed points, we have the fixed points $g_1=h_1=\epsilon/(N+4)$, and 
$(g_2,h_2)$ with $g_2=N\epsilon/(N^2+8)$ and $h_2=(4-N)\epsilon/(N^2+8)$. From these only 
the fixed point $(g_2,h_2)$ is infrared stable, provided $N=3$. Note that for $N=4$ 
we have the realization of the deep easy-plane limit, since 
the effective interaction multiplying $|z_1|^2|z_2|^2$ vanishes, although 
the bare coupling $w_0\neq 0$. However, such a 
case does not corresponds to a stable fixed point. 
Note that for $N=2$ the fixed point $(g_2,h_2)$ is 
$O(4)$ symmetric and coincides with $(g_1,h_1)$.  

Note that the gauge field fluctuations, which in this problem are essential, 
generate a $|z_1|^2|z_2|^2$ term. Thus, we  have to keep $w_0\neq 0$ and look  
in the RG treatment for the stability of fixed points with $h=0$.    
It turns out that {\it for all values of 
$N$, no fixed points with $h_*=0$ and $f_*=3\epsilon/N$ are found}. 
Therefore, no second-order phase transition takes place in this case. 
This is a very significant result, since as we have discussed, the 
existence of a critical value of $N$ above which the transition 
becomes second-order reflects the actual behavior at lower values of $N$. 
The complete absence of fixed points for all $N$ provides a solid 
theoretical explanation for the numerical results of Refs. \cite{Kuklov_2006} 
and \cite{Kragset}. 
Fixed points with $h_*\neq 0$ exist for large enough $N$, but none of them are stable.  
Therefore, there is no deconfined quantum criticality associated to the 
model (\ref{aniscp1}). 

Summarizing, we have considered three models for deconfined spinons.  
From the three models considered, only the one associated with an easy-plane antiferromagnet 
does not exhibit any second-order phase transition, in 
agreement with the numerical results of Refs. \cite{Kragset} and \cite{Kuklov_2006}. 
Deconfined spinons were shown to govern 
a second-order phase transition for both the isotropic $SU(N)$ antiferromagnet and quantum spin nematic 
systems. In both cases we have computed the critical exponent $\eta$ using the scaling relation 
$\eta=d+2(1-\varphi/\nu)$ in terms of the crossover exponent $\varphi$ and the correlation 
length exponent $\nu$. Knowledge of the 
exponent $\varphi$ is very important in the study of crossover behavior and 
stability of frustrated systems.       

The authors thank Z. Tesanovic for enlightening discussions, and the Centre for Advanced 
Studies at the Norwegian Academy of Sciences and Letters, for hospitality and financial 
support. A.S. thanks the Freie Universit{\"a}t Berlin for hospitality. This work was 
supported by the Research Council of Norway, Grants No. 157798/432 and No. 158547/431 
(NANOMAT), and Grant No. 167498/V30 (STORFORSK)


\begin{thebibliography}{100}
\bibitem{Note} The critical exponent $\eta$ discussed here should not be confused with the one 
calculated in Ref. \cite{Halperin-Lubensky-Ma} in the context of $N$-component superconductors. 
There $\eta$ is the exponent of the superconducting order parameter, i.e., it is 
associated with a two-field correlation function and {\it is not} gauge-invariant.
\end{thebibliography}
\end{document}